\documentclass[a4paper,11pt]{article}

\title{Breaking the uniqueness of the Shape Dynamics Hamiltonian}
\author{\bf Henrique Gomes\footnote{\href{mailto:gomes.ha@gmail.com}{gomes.ha@gmail.com}}\\\it  Department of Physics,  University of California, Davis,   CA, 95616}

\usepackage{amssymb}
\usepackage{amsmath}
\usepackage[usenames,dvipsnames]{color}
\usepackage{hyperref}
\usepackage[left=1in,right=1in,top=1in,bottom=1in]{geometry}                  % For good copy.

\hypersetup{
    colorlinks=true,         % false: boxed links; true: colored links, false is default
    linkcolor=blue,          % color of internal links, red is default
    citecolor=red,        % color of links to bibliography, 'green' is default
    urlcolor=Violet             % color of external links, cyan is default
}

\let\oldmarginpar\marginpar
\renewcommand\marginpar[1]{\oldmarginpar{\color{red}\raggedright\scriptsize #1}}

\newcommand{\mean}[1]{\ensuremath{\lf\langle #1 \rt\rangle }}
\newtheorem{proposition}{Proposition}

\newtheorem{theorem}{Theorem}

\def\be{\begin{equation}}
\def\ee{\end{equation}}
\def\bea{\begin{eqnarray}}
\def\eea{\end{eqnarray}}

\def\lf {\ensuremath{\left}}
\def\rt {\ensuremath{\right}}

\begin{document}

\maketitle
\begin{abstract}
In \cite{Gomes:2011zi} a linear method of solving a particular set of Lichnerowicz-type equations through the implicit function theorem was sketched  in order to implicitly construct Shape Dynamics' global Hamiltonian and eliminate  second class constraints. This method was completely laid out in  \cite{Gomes:thesis},  and in \cite{Gomes:2011au} it was used  for extending Shape Dynamics (SD) to the non-vacuum case,  showing how other fields are coupled to the theory. In the latter paper it was noticed that, unlike the vacuum case,  the use of such methods yielded puzzling bounds on the density of some types of fields. Here we show that the original SD cannot be extended beyond such bounds, but that a slight modification of the original can withstand any type of coupling. When the bound is broken, the theory does not come equipped with a single Hamiltonian as in vacuum SD, but with a finite set of weakly commuting Hamiltonians, which we describe. 
\end{abstract}

\section{Introduction}
\subsection{Shape Dynamics in a nutshell}
Let us start by  very briefly defining the setting in which this study takes place: the theory of Shape Dynamics (for a more comprehensive introduction see \cite{Gomes:2010fh, Gomes:2011zi, Gomes:thesis}).\footnote{For the original motivation to regard the role of conformal transformations in ADM, see \cite{Anderson:2004bq}.} In very brief terms, SD is a classical theory equivalent to pure gravity in which the spacetime picture that underlies GR is replaced by a picture of evolving 3-geometries, as in the original intention of Wheeler. What makes SD unique amongst more standard geometrodynamical theories of gravity, is that the usual  invariance under spacetime refoliations is traded for invariance of the theory  under local spatial conformal transformations that preserve the total spatial volume. The equivalence is obtained by manipulating the constraint structure of General relativity, but can only be achieved if one first uses the Stuckelberg mechanism to extend the original  phase space with extra degrees of freedom. After  this ``Stuckelberg extension" of the original ADM dynamical system, one obtains a system even more redundant in symmetries,\footnote{In fact, the most technically clear way of obtaining SD requires first of all a trivial embedding of the original system into the extended phase space, and only then a canonical transformation that in effect introduces a conformal character to the system.}   which is reducible to both the original ADM and SD through the use of different gauge conditions on the added variables. 

 The basic condition required for the symmetry trading is the existence of two (the original and the ``new") symmetries which are maximally symplectic (i.e. gauge fix one another). Although this basic condition can be expressed in the original ADM phase space, the construction of SD itself requires us to use extended phase space explicitly. It is then the  implicit function theorem which enables the use of a linear method and the maximally symplectic character of the dual symmetries to obtain SD from extended phase space. This linear method, fully laid out in section 4.3  of \cite{Gomes:thesis}, works perfectly well in vacuum SD. It shows that \emph{any} ADM vacuum dynamical system has its SD counterpart, provided we interpret SD in a purely geometrodynamical way, as explained here in section \ref{sec:tech_vacuum}. However, as the author found out,  the original linear method was not comprehensive enough to deal with certain non-vacuum systems. This result was  first reported in \cite{Gomes:2011zi}. It is with the extension of this method that the present paper is concerned.

\subsection{Non-uniqueness of the SD Hamiltonian}

The original SD construction allowed the linear method to reduce an extended theory to a theory existing on the ADM phase space, possessing a single global Hamiltonian and local conformal invariance. It relied heavily however on the positivity of the linear term in a differential operator arising upon reduction of the system. Unfortunately, once one tries to extend the result to couple matter, the required term is no longer positive, and thus the relevant differential operator may have non-trivial homogeneous solutions. Conceivably, these could have yielded different theories of Shape Dynamics, or perhaps the reduced system could have turned out to be inconsistent. 

The present paper will use ellipticity of the relevant differential operators to extend that result and show that in the general case one still retains local conformal invariance, but must allow for the possibility of a finite set of commuting global Hamiltonians instead of the single one. The number of commuting global Hamiltonians is related to the kernel of the relevant differential operator. Here we show that there is still a coherent ``Shape Dynamics" theory arising, at least for the case of a compact closed 3 manifold.

Before we start, we give a warning about the level of mathematical rigor aimed for in this paper. Although the paper contains proofs and propositions, it is aimed at a physics audience. Its purpose is to arrive at a new \emph{physical} result regarding SD, namely that for some types of matter (including certain values of the cosmological constant) we have a theory dynamically equivalent to ADM with a finite number of compatible notions of ``absolute" (global) times.  We are \emph{not} extremely concerned with the type of functional spaces we will be dealing with, and often sloppily characterize distributions as elements of $C^\infty(M)$, and other such mathematical atrocities. However, we have aimed to give robust, plausible arguments, such that if due care was given to such domains all statements here could be put in a firm functional analytic grounding.\footnote{For example one could use Sobolev completions to given finite rank Sobolev spaces and then inverse limits to attain such rigour. Or simply work with $C^2(M)$ instead of $C^\infty(M)$. } We will usually assume that we are dealing with Hilbert spaces endowed with an $L_2$ inner product.

\section{The vacuum case}

We now give a more technical, but very streamlined account of the construction of SD. This will be useful to understand the limitations imposed by the bound, and the attempt to overcome these later. 

\subsection{Technical setting of SD}

Let us now briefly review the setting for the construction of Shape Dynamics as a theory equivalent to ADM gravity on a compact Cauchy surface $\Sigma$ without boundary.  We start with the standard ADM phase space $\Gamma_{\mbox{\tiny ADM}}=\{(g,\pi):g\in \mathrm{Riem},\pi\in T_g^*(\mathrm{Riem})\}$, where $\mathrm{Riem}$ denotes the set of Riemannian metrics on the above defined 3-manifold $\Sigma$, and the usual first class ADM constraints, i.e. the scalar constraints
\be S(x)=\frac{\pi^{ab}\pi_{ab}-\frac{1}{2}\pi^2}{\sqrt g}-\sqrt g R\ee
 and momentum constraints
\be H^a(x)=\pi^{ab}_{~;b}(x)\ee thereon. We extend the ADM phase space with the phase space of a scalar field $\phi(x)$ and its canonically conjugate momentum density $\pi_\phi(x)$, which we introduce as additional first class constraints $\mathcal{Q}(x)=\pi_\phi(x)\approx 0$. The system is thus merely a trivial embedding of the original ADM onto the extended phase space $\Gamma_{\mbox{\tiny ADM}}\times \Gamma_\phi$ and no sign of a ``conformal transformation" is in sight. 

It is by introducing the canonical transformation $ \mathcal {T}_{\phi}$ generated by the generating functional $F=\int d^3x\left(g_{ab}e^{4\hat \phi}\Pi^{ab}+\phi\Pi_\phi\right)$,  where $\hat \phi(x):=\phi(x)-\frac 1 6 \ln\langle e^{6\phi}\rangle_g$ using the mean $\langle f\rangle_g:=\frac 1 V \int d^3x\sqrt{|g|} f(x)$ and 3-volume $V_g:=\int d^3x\sqrt{|g|}$ that a conformal character of the theory starts to appear. For it is this transformation that emulates a volume-preserving conformal transformation in the original canonical variables:
\begin{equation}\label{equ:canonicalTransformation_grav}
 \begin{array}{rcl}
   g_{ab}(x)&\to& \mathcal {T}_{\phi} g_{ab}(x):=e^{4\hat\phi(x)}g_{ab}(x)\\
   \pi^{ab}(x)&\to&\mathcal {T}_{\phi} \pi^{ab}(x):=e^{-4\hat\phi(x)}\left(\pi^{ab}(x) -\frac{g^{ab}}{3}\sqrt {g}\langle \pi\rangle (1-e^{6\hat\phi})\right)\\
      \phi(x)&\to&\mathcal {T}_{\phi} \phi(x):=\phi(x)\\
   \pi_\phi(x)&\to&\mathcal {T}_{\phi} \pi_\phi(x):=\pi_\phi(x)-4(\pi(x)-\mean{\pi}\sqrt g).
 \end{array}
\end{equation}
Renaming for convenience:
$$4(\pi(x)-\mean{\pi}\sqrt g=:D(x)
$$ at this point we have the first class set of constraints
\be   \mathcal {T}_{\phi} S(x),~
    \mathcal {T}_{\phi} H(x)~\mbox{ and}~
    \mathcal {T}_{\phi} Q(x)=\pi_\phi(x)-D(x)\ee We should note however that the extended system does not yet possess any notion of ``conformal symmetry", for the  action of $ \mathcal {T}_{\phi} Q(x)$ on the canonical variables $\phi$ and $\pi_\phi$ is not conformal. We also note that $ \mathcal {T}_{\phi} Q(x)$ acts trivially on phase space functionals of the form $ \mathcal {T}_{\phi}( f(g,\pi)(x))$ and that $\mathcal {T}_{\phi} H(x)$ acts (weakly) as diffeomorphisms in extended phase space.
 
It is the presence of the volume-preserving element of the conformal transformations that accounts for the more complicated structure of the momentum transformation, and it is also the source for the presence of a non-trivial global Hamiltonian in SD, as we show now.

 To regain ADM, one imposes the gauge-fixing $\phi=0$. To obtain SD, one introduces into the extended system above the gauge-fixing $\pi_\phi=0$. To get SD from the imposition of this gauge-fixing is where it is most convenient to use the linear method, as we now explain.

\subsection{Finding and solving the second class constraints}
Again, this is a streamlined version of the entire argument. For a more comprehensive and in-depth treatment, see \cite{Gomes:thesis}, section 4.3, where the method was first completely laid out. As a matter of nomenclature, we say that a system of constraints is first class if it commutes on the constraint surface (weakly), and it will be ``purely" second class, if the Poisson brackets are invertible (for example, if they are proportional to the identity). If this is the case the bracket itself cannot  impose  further restrictions on phase space (constraints), and the second class constraints must be either completely solved for (used as definitions of given canonical variables), or one must use the Dirac bracket to project the symplectic structure onto the constraint surface. We will show that the first case is attainable here in an elegant manner which gets rid of extra variables.  

\subsubsection{Separate out the first class component}

The only weakly non-vanishing Poisson-bracket of the gauge-fixing condition $\pi_\phi(x)=0$ with the constraints of the linking theory is
\be\label{equ:Pb_pi_phi,pi}\{\overline{{\mathcal {T}_\phi}S}(N),\pi_\phi(x)\}=4\mathcal {T}_\phi\{S(N),\pi(x)-\mean{\pi}\sqrt g(x)\},\ee
By the canonical transformation properties of the theory, we can translate any assertion made at $\phi=0$ to a different $\phi$ by the use of $\mathcal {T}_\phi$. Hence the relevant Poisson bracket becomes:
\begin{equation}\label{lapse}4\{S(N),\pi(x)-\mean{\pi}\sqrt g(x)\}=8\Delta N(x)\sqrt g-8\langle \Delta N\rangle\sqrt g-6S(x)N(x).\end{equation}
   In \eqref{lapse} $\Delta$ is the differential operator:
\begin{equation}
\label{Delta}\Delta=\nabla^2-\frac{1}{4\sqrt g}\langle\pi\rangle\pi-R.
\end{equation}
On the gauge fixed constraint surface $\mathcal {T}_{\phi} S(x)= \mathcal {T}_{\phi} Q(x) =\pi_\phi=0$ at $\phi=0$,\footnote{We have taken all the relevant brackets, so now it is legal to regard the behavior of the operator on the constraint surface, and at $\phi=0$, and then generalize by using the canonical transformation properties (which preserve the brackets). } using  the momentum split $\pi^{ab}=\bar\sigma^{ab}\sqrt g+\frac{1}{3}g^{ab}\pi$ into its trace and  traceless part, $\sigma^{ab}=\bar\sigma^{ab}\sqrt g$, we have
\be\label{equ:new_Delta}\Delta\approx\nabla^2-\frac{1}{12}\langle\pi\rangle^2-\bar\sigma^{ab}\bar\sigma_{ab}.
\ee
Thus our operator $\Delta$ can be written as $\Delta=\nabla^2-A$ where
  $A[g,\pi;x):=\bar\sigma^{ab}\bar\sigma_{ab}+\frac{1}{12}\mean{\pi}^2\geq 0$, i.e. it is a positive-definite function, vanishing only when $\pi^{ab}=0$. Rewriting \eqref{lapse} in a more convenient manner we obtain:
\be\label{lapse2}\frac{1}{\sqrt g}\{S(N),\pi(x)-\mean{\pi}\sqrt g(x)\}\approx2(\Delta N-\mean {\Delta N})
\ee

 The first question to be asked now is: are the given constraints purely second class? I.e. is the Poisson bracket matrix \eqref{lapse} invertible? The answer is dependent on the existence of  homogeneous solutions, $\tilde N(x)\not\equiv 0$ such that:
\be\label{equ:Delta_constr} \Delta \tilde N=\mean {\Delta \tilde N}
\ee
If there is such a solution,  the answer is no,  the bracket is not invertible as it stands. 

 The initial challenge then is to try to separate the constraints $\mathcal {T}_{\phi} S$  into a purely second class part, which thus has invertible Poisson brackets with $\pi_\phi=0$, and is thus completely gauge-fixed by this condition, and another part, which weakly commutes with the gauge-fixing condition. 

From the positivity of the linear term $A$ in $\Delta$, we understand that it cannot have a non-trivial homogeneous solution on a compact manifold, i.e. $\Delta N=0$ implies $N=0$. This follows from comparing both sides of $\Delta N=0$ at the maximum (minimum) of $N$. An equivalent statement is that for any function $f$ there exists a unique $N_f$ such that $\Delta N=f$.  Furthermore, a solution of \eqref{equ:Delta_constr} has to be such that $\Delta N_0=c$, where $c$ is a spatial constant. Since any constant will do, we have a linear space of solutions at each time, as it should be. We fix a generator of this space by adjusting the constant $c$ to be such that $\mean {N_0}=1$ which is guaranteed to be possible, since for $c<0$, $N(x)\geq 0$ for all $x$. The truth of this last statement is apparent  in the following manner: suppose that $N(x)<0$ for some $x$, then for some $y$,  $N(y)$ will attain a minimum, and thus $\nabla^2N(y)>0$. The equation then becomes $A=-\nabla^2 N(y)-|c|$, which is absurd. Note that almost all of the assertions we have made so far require positivity of $A$, and it is exactly this positivity that will be broken in section \ref{sec:bound}. 

\subsubsection{Using the linear method to solve the second class parts.}\label{sec:tech_vacuum}

We have succeeded in separating out one generator of the first class part of $\mathcal{T}_\phi S$ with respect to $\pi_\phi$, namely $\mathcal{T}_\phi (S(N_0))$, where $N_0$ obeys \eqref{equ:Delta_constr} and $\mean{N_0}=1$. We have not yet singled out the uniquely second class part, which has invertible Poisson bracket wrt $\pi_\phi$. After we have done that, to arrive at Shape Dynamics, we still must show that that second class part is solvable by a definition of the conjugate variable to $\pi_\phi$, i.e. that we can choose $\phi=\phi_o(g,\pi)$ such that this solves the second class component of $\mathcal{T}_\phi S$. We will then have the complete theory on the original phase space of ADM, with constraints  $\mathcal{T}_{\phi_o} (S(N_0))~,~ D~, ~H$. It is here that we will use the implicit function theorem.

First, it will prove very useful to be more careful about what we mean with ``a smearing".  Consider
 \begin{equation}\label{equ:def}{ \overline{\mathcal{T}S}(x)}:\Gamma\times T^*(C^\infty(M)/\mathcal{V})\rightarrow C^\infty(M),\end{equation}  
where $\mathcal{V}$ is the linear space of non-zero constant functions and the overline means we take the de-densitized version. I.e., suppose we have some density: $F:\Gamma\rightarrow D(M)$, where $D(M)$ is the space of densities of weight one over $M$. Then $\bar F(g,\pi)=F(g,\pi)/\sqrt g$. \footnote{We again emphasize that these statements could be made rigorous using the constraints to be taken as test functions and the smearings as the space of linear functionals over $C^\infty(M)$. This would resolve some issues to come as for example the use of the delta function as a smearing. }

 Since the map \eqref{equ:def} does not depend on $\pi_\phi$, we can fix $\pi_\phi(x)=f(x)$. Then
  \begin{equation}\overline { {\mathcal {T}}S}(x)_{\pi_\phi=f(x)}:\Gamma\times C^\infty(M)/\mathcal{V}\rightarrow C^\infty(M).
\end{equation}
 We note that in fact ${\mathcal {T}_\phi}S(x)$ depends solely on $C^\infty(M)/\mathcal{V}$. In any case, a smearing is given once we have established the $L_2$ inner product:
\be\langle \bar F, h\rangle:=\int_\Sigma \bar F h \sqrt g d^3x =\int_\Sigma  F h d^3x =:\langle F, h\rangle
\ee
This creates a pairing between the space of densities and that of functions, which we defined in the above rhs.

Consider the linear operator:
   \be\label{equ:tangent}\delta_{\mathcal{C}}\overline{{\mathcal {T}_\phi}S}(g_0,\pi_0,\pi_\phi^0)_{|\phi=0}: {T}_0(C^\infty(M)/\mathcal{V})\rightarrow C^\infty(M)\ee
  Here we have denoted the derivative in the second coordinate, the one parametrized by $\phi$, by a subscript $\mathcal{C}$, and  where $T_x\mathcal {N}$ denotes the tangent space to $\mathcal{N}$ at $x\in \mathcal{N}$, and, as in usual partial derivatives, one holds the coordinates $(g,\pi,\pi_\phi)$ fixed. We will omit from now on the ``initial" point $(g_0,\pi_0,\pi_\phi^0)$ where we take the derivative. It can be checked that the space ${T}_0(C^\infty(M)/\mathcal{V})$ can be redundantly parametrized by functions $f\in C^\infty(M)$ as $f-\mean f$, so that constant functions are mapped to the zero element in ${T}_0(C^\infty(M)/\mathcal{V})$.

The tangent map is given by:
\begin{equation}\label{iso2} \delta_{\mathcal{C}}{{\mathcal {T}_\phi}S}_{|\phi=0}:=\frac{\delta {{\mathcal {T}_\phi}\mathcal{S}(x)}}{\delta\phi(y)}_{|\phi=0}=\{{\mathcal {T}_\phi}S(x),\pi_\phi(y)\}_{|\phi=0}
=\sqrt g(x)(\Delta(x)\delta(x,y)-\frac{\Delta(x)}{V})\end{equation}
 Contraction of \eqref{iso2} with $N(x)$ yields  $\Delta N-\langle \Delta N\rangle$. We note that contraction in the $x$ variable requires us to use the adjoint $\delta_{\mathcal{C}}{\mathcal {T}_\phi}S^*_{|\phi=0}$, and as this is \emph{not} a self-adjoint operator the distinction is important. Thus
 \be\label{equ:tangent2}(\delta_{\mathcal{C}} \overline{\mathcal {T}_\phi S})^*\cdot N= \Delta N-\langle \Delta N\rangle
 \ee
and from  the uniqueness of the kernel of the adjoint we in fact have 
\be \mbox{Im}(\delta_{\mathcal{C}} \overline{\mathcal {T}_\phi S})\oplus N_0=C^\infty(M)
\ee

 Now we construct a modification of $ S(x)$, which we will call $L$ such that it has the same tangent map but  its range must be such that:
  \begin{equation}({\mathcal{T}_\phi L})_{\pi_\phi=f(x)}:\Gamma\times C^\infty(M)/\mathcal{V}\rightarrow \text{Im}(\delta_{\mathcal{C}} \mathcal {T}_\phi S).
\end{equation}
 This map is given by  $L= S(x)- S(N_0)\sqrt{g}$, since as one can readily check the tangent map $\delta_{\mathcal{C}} {\mathcal {T}_\phi L}$  indeed stays (weakly)  the same and 
 $\langle L, N_0\rangle=0$ (since $\mean {N_0}=1$).

We have \emph{not} yet proven that $\delta_{\mathcal{C}}\overline{\mathcal {T}_\phi L} $ is a topological linear isomorphism (or alternatively that $\mathcal {T}_\phi L$ is purely second class with respect to $\pi_\phi$).  We have shown that it is a surjective linear map (or alternatively that $\mbox{Ker}(\delta_{\mathcal{C}} \overline{\mathcal {T}_\phi L})^*=0$), but we must still prove injectivity, i.e. that  $\mbox{Im}(\delta_{\mathcal{C}} \overline{\mathcal {T}_\phi L})^*= T_0(C^\infty(M)/\mathcal{V})$. As we know, the elements of the latter space are given by $f-\mean {f}$ where $f\in C^\infty(M)$. 
The differential operator $\Delta$ is invertible, possessing a Green's function. Thus for any function $f$ there exists a unique $N_f$ for which $\Delta N_f=f$. Thus $(\delta_{\mathcal{C}} {\mathcal {T}}_\phi S)^*\cdot N_f=\Delta N_f-\mean{\Delta N_f}=f-\mean f$, and we have surjectivity of the formal adjoint. Let us here briefly show an alternative proof of injectivity, which will be more useful in the generalization of the following section. First, we have that since we are assuming a Hilbert space structure,  $\mbox{Ker}(\delta_{\mathcal{C}} \overline{\mathcal {T}_\phi L})=(\mbox{Im}(\delta_{\mathcal{C}} \overline{\mathcal {T}_\phi L})^*)^\perp$. Thus, by the invertibility properties of $\Delta$, any element of  $(\mbox{Im}(\delta_{\mathcal{C}} \overline{\mathcal {T}_\phi L})^*)^\perp$ must satisfy  $\langle f-\mean{f},\rho_0\rangle=0$ for any  function $f$.

Suppose then that $\rho_0(y)\neq 0$ for some   $y\in M$. Let us take $f_y(x)=\delta(x,y)\rho_0(x)$ (i.e. we take the point source of the field $\rho_0$). Then $f_y(x)-\mean f_y=\delta(x,y)\rho(x)-\frac{\rho(y)}{V}$ and
\be\label{equ:injectivity1}\langle f_y(x)-\mean{f_y},\rho_0\rangle=\rho_0^2(y)-\rho_0(y)\mean{\rho_0}=0\ee
which means $\rho_0(y)=\mean {\rho_0}$, and thus the equivalence class of $\rho$ in $T_0(C^\infty(M)/\mathcal{V})$ is zero, which makes the map injective. 

 By the canonical transformation properties of $\mathcal{T}_\phi$, one can extend this construction to arbitrary $\phi$. We have thus proven
\begin{proposition}\label{prop2}
The linear map given by $\delta_{\mathcal{C}}\overline{\mathcal {T}_\phi L}(x):T_0(C^\infty(M)/\mathcal{V})\rightarrow \text{Im}(\delta_{\mathcal{C}}\overline{ \mathcal {T}_\phi S})\simeq C^\infty(M)/N_0$
where ${\mathcal {T}_\phi L}(x)=\mathcal {T}_\phi S(x)-T_\phi( S(N_0))\sqrt{g} $, is a toplinear isomorphism for all $(\phi, g,\pi)$ provided $\pi^{ab}\not\equiv 0$.
  \end{proposition}
  We have shown that it is a linear continuous bijection, and hence a topological linear isomorphism. $\square$.  

This singles out the equations we must solve for $\phi$, namely, $\mathcal {T}_\phi L=0$. But we can go further, not only can we form the Dirac bracket using $\{{{\mathcal {T}}_\phi L}(x),\pi_\phi(y)\}^{-1}$,  but we can now use the implicit function theorem for Banach spaces for the function $\overline{\mathcal {T}_\phi L}_{\pi_\phi=f(x)}:\Gamma\times \mathcal{C}/\mathcal{V}\rightarrow \text{Im}(\delta_{\mathcal{C}}\overline{ \mathcal {T}_\phi S})\simeq C^\infty(M)/N_0 $. First we remind the reader of the formulation of the implicit function theorem for Banach spaces: 
\begin{theorem}[Implicit function]
Let $X, Y, Z$ be Banach spaces. Let the mapping  $f:X\times Y\rightarrow Z$ be continuously differentiable. If  $(x_0,y_0)\in X\times Y$, such that $f(x_0,y_0)=0$ and $\delta_Yf_{(x_0,y_0)}:Y\rightarrow Z$ is a Banach space isomorphism from $Y$ onto $Z$, then there exist neighborhoods $U$ of $x_0$ and $V$ of $y_0$ and a differentiable function $g:U\subset X\rightarrow V\subset Y$ such that $f(x,g(x)) = 0$ and $f(x,y) = 0$ if and only if $y = g(x)$, for all $(x,y)\in U\times V$.
\end{theorem}

In the above theorem, we substitute $X\mapsto \Gamma, Y\mapsto C^\infty(M)/\mathcal{V}, Z\mapsto C^\infty(M)/N_0$, and $f\mapsto \mathcal{T}_\phi L$.  Now, for any $(x_0,y_0)$ on the constraint surface $\{\mathcal{T}_\phi S=0\}\cap \{D=0\}$ (which also implies that $\mathcal{T}_{y_0}( L(x_0))=0$) we have that:
\begin{theorem}\label{prop}
 Around each point on the constraint surface there exist open subsets $U\times V\in \Gamma\times C^\infty(M)/\mathcal{V}$ such that there  exists a unique $\hat\phi_0:U\rightarrow \mathcal{C}/\mathcal{V}$,   such that the level surface formed from ${\mathcal {T}_\phi L}=0$ is given by  \end{theorem}
 $$\{(g_{ij},\pi^{ij},
 \hat\phi_0[g_{ij},\pi^{ij}],\pi_\phi)\in U\times (\Gamma_\phi)_{V}\}.
$$ In other words, around each point on the constraint surface we can find the solution to ${{\mathcal {T}_\phi L}}(g,\pi,\phi,\pi_\phi)=0$ for all $(g,\pi,\pi_\phi)$ in a given open neighborhood  by setting $\phi=\phi_0$.\footnote{Different solutions might exist around different open neighbourhoods, but we will not enquire on how these can be glued together.} Furthermore, since we have in the process shown that $\delta_{\mathcal{C}}\overline{ \mathcal {T}_\phi L}$ is surjective, the level surface formed by $({\mathcal {T}_\phi L})^{-1}(0)$ is regular. 
$\square$.

Thus we reach the surprising result that we can \emph{solve the second class constraints while eliminating the extra Stuckelberg variables!} Although using the implicit function theorem in this context seems a bit contrived, we have a powerful theorem that is guaranteed to work whenever the operator $\Delta$ given in \eqref{Delta} is invertible. Although the operator $\Delta$ concerns only functionals in the original phase space $\Gamma$, the canonical transformation in the extended phase space not only tells us we can extend this result, but also that it will tell us a lot about the tangent of these maps on the $\phi$ direction, enabling us in the end to use the implicit function theorem.

We end this section by noting three things. First of all, the emerging theory, with constraints $\mathcal{T}_{\phi_o} (S(N_0))~,~ D~, ~H$ is clearly not merely ADM in a CMC foliation, for it contains the extra, conformal constraint $D$, and the map $\mathcal{T}_{\phi_o}$, which do not appear in ADM under any guise. Secondly, we note that since $N_0$ is not identically zero, the theory makes sense as a theory of geometrodynamics even if $N_0(x)=0$ for some $x\in \Sigma$. This is why we have asserted that \emph{any} ADM dynamical system has an SD counterpart, even if  some vacuum  ADM spacetimes do not  have a CMC foiliation. Space-times not admitting a CMC foiliation are taken to be those where the lapse solution $N_0$ reaches zero at some point, thus freezing time. In the geometrodynamical sense which we adopt here this assertion loses meaning. Lastly, we note that we did not solve the Lichnerowicz-York equation $S(e^{4\phi}g, e^{4\phi}\pi)=0$, where the $\phi$'s are not restricted to maintain volume. In fact, we have not even solved our  version ${\mathcal {T}_\phi S}=0$, but only the non-homogeneous version ${\mathcal {T}_\phi L}=0$ allowing a global Hamiltonian to be left after reduction. 

In this direction we also note that one could possibly use the York-\'O'Murchada method of \cite{York-Niall} to solve such types of equations, but the criteria for solvability becomes more complicated, and much harder to use. In \cite{Gomes:2011au}, the authors originally set out to use both methods in parallel, but the flexibility and simplicity of the linear method in the end convinced us to leave the original York-\'O'Murchada method out. Furthermore, even if we were to use said methods,  we must first come to the clean separation between first and second class constraints. I.e. we must come at least as far in our analysis as proposition \ref{prop2} to know \emph{what} is the equation we would like to apply a given method  to, namely, what is the purely second class constraint in question. The advantage of our method is that from that point on it requires no extra work. 

\section{Breaking the bound}\label{sec:bound}

Now we mention briefly what becomes of our method once we couple different types of fields to the original construction. We will find that the equivalent resulting linear part of $\Delta$ is no longer positive, and we will find ways to circumvent this in the construction of SD.  We have outlined the whole method of construction in the previous section  so that we can follow a similar pattern for the more general case presented now. 
\subsection{Coupling of matter fields}
The coupling of different types of matter fields was done more completely in \cite{Gomes:2011zi}, and reported more briefly in \cite{Gomes:loops}. It was there found that if we assume an independent scaling of fields as in, for example in the scalar case, $\psi\rightarrow e^{n\hat\phi}\psi$, for $\alpha\neq 0$ the propagation of gauge constraints develops pathologies which the authors could not amend. For this and other reasons, it was found that the natural choice for fields was ``neutral coupling". 
 Neutral coupling is the particular choice of \emph{empty} scaling $\alpha=0$, i.e. non-gravitational fields have conformal weight zero. This means that fields are only scaled in the sense that they are ``carried along" by the scaling of the spatial metric. The method then proceeds as  before, as we now briefly describe. 

We start with the general, first class  constraints
\begin{equation}
 \begin{array}{rcl}
   H(\xi)&=&\int d^3x \left( \pi^{ab} (\mathcal L_\xi g)_{ab}+ \pi^A \mathcal (L_\xi \psi_A) \right)\\
   S(N)&=&\int d^3x\left(\frac 1{\sqrt{|g|}}\pi^{ab}G_{abcd}\pi^{cd}-(R-2\Lambda)\sqrt{|g|}+S_{\mbox{\tiny matter}}(g_{ab},\psi_A,\pi^A)\right)N(x)\\
G^\alpha(\lambda_\alpha)&=&\int d^3x G^\alpha(g_{ab},\psi_A)\lambda_\alpha\sqrt g
 \end{array}
\end{equation}
where we denote the ``matter" (or rather, just non-gravitational) degrees of freedom collectively by $\psi_A$ and their canonically conjugate momenta $\pi^A$. We assume that the matter Hamiltonian $H_{\mbox{\tiny matter}}$ does neither contain $\pi^{ab}$ nor any spatial derivatives of $g_{ab}$. This is a reasonable assumption, and it holds for all fields known. We furthermore assume the constraint associated to internal gauge symmetries to be a functional of only the ``position" variables $\psi^A, g_{ab}$. This is also a condition realized in all matter fields studied, and it simplifies treatment greatly. 

The Stuckelberg extension proceeds as before, but now we thus use the canonical transformation generated by:
\begin{equation}
 F=\int d^3x\left(e^{4\hat \phi} g_{ab}\Pi^{ab}+\phi \Pi_\phi+\phi_A \Pi^A\right)
\end{equation}
which  acts non-trivially only on the gravitational variables, as required by neutral coupling. The resulting canonical transformations leaves the extra constraint $\mathcal{T}_\phi Q$ as in the vacuum case,  the transformed diffeomorphism constraint $H$ still weakly generates diffeomorphisms in the extended phase space (now also extended by the matter degrees of freedom), and the  gauge constraint decouples from the conformal transformation, so that $T_\phi G^\alpha\propto G^\alpha\approx 0$.

After imposing the gauge-fixing $\pi_\phi(x)=0$ we always get an equation of the form:
\be  \{T_\phi S(N), \pi_\phi(x)\}=T_\phi\left[-\frac{3}{2}S(x)+2\sqrt g( \Delta_{\mbox{\tiny matter}}N(x)-\mean{ \Delta_{\mbox{\tiny matter}}N})\right]
\ee
If we here restrict to matter Hamiltonians which don't n contain spatial derivatives of the metric tensor nor metric momenta, we get:
\be\label{equ:def:Delta_gen}\Delta_{\mbox{\tiny matter}}:= \nabla^2-\frac{\pi\langle\pi\rangle}{4\sqrt g}-R+\frac{1}{2\sqrt{|g|}}\left(\frac{\delta S_{\mbox{\tiny matter}}}{\delta g_{ab}}g_{ab}+\frac{3}{2}S_{\mbox{\tiny matter}}\right)\ee
where we note a slight deviation from previous notation, since here the subscript ``matter"  on the $\Delta$ operator serves to make it distinct from the vacuum case, and  does not refer somehow only to the operator regarding the matter degrees of freedom. On the constraint surface $T_\phi S=0$ and $\mathcal{Q}=0$, the end result is equivalent to taking
\be\label{equ:Delta_matter} \Delta_{\mbox{\tiny matter}}\approx (\nabla^2-\frac{1}{12}\langle\pi\rangle^2)-\frac{\sigma^{ab}\sigma_{ab}}{g} 
+\frac{1}{2\sqrt{g}}\left(\frac{\delta S_{\mbox{\tiny matter}}}{\delta g_{ab}}g_{ab}-\frac{1}{2}S_{\mbox{\tiny matter}}\right)\ee
Thus in this case the criterium for invertibility of the operator rests on:
\be\label{equ:inv_criterium}\frac{1}{2\sqrt{g}}\left( \frac{\delta S_{\mbox{\tiny matter}}}{\delta g_{ab}}g_{ab}-\frac{1}{2}S_{\mbox{\tiny matter}}\right)\leq  \sqrt{g} \frac{1}{12}\langle\pi\rangle^2 +\frac{\sigma^{ab}\sigma_{ab}}{ g} 
\ee
the so called ``bound"  found  in \cite{Gomes:2011au}. So the question that we face is: what happens if the bound is broken?

\subsection{Finding the second class parts in the general case}

\subsubsection{Properties of the solutions $N_i$}
Restricting $S_{\mbox{\tiny matter}}$ to not contain any metric derivatives, guarantees us that the symbol of the operator $\Delta_{\mbox{\tiny matter}}$ given in \eqref{equ:Delta_matter} is defined by $\nabla^2$ and thus $\Delta_{\mbox{\tiny matter}}$ is elliptic. 
Taken to be  an operator between Sobolev spaces, $\Delta_{\mbox{\tiny matter}}$ is then a \emph {Fredholm} operator, and as such possesses the very important property \emph{that it has a finite dimensional kernel and cokernel} \cite{Nash}, where  the cokernel of a linear mapping of vector spaces $f : X \rightarrow Y$ is the quotient space $ Y/im(f)$ of the codomain of $f$ by the image of $f$. This immediately guarantees that such operators \emph{are invertible modulo compact operators}. These are the technical facts that allows us to have a similar construction as in the invertible case. So let us follow the constructions from the previous section. 

First, to obtain the isomorphism we will still have to study the analogous equation to \eqref{equ:Delta_constr}
\be\label{equ:Delta_constr_matter} \Delta_{\mbox{\tiny matter}} \tilde N=\mean {\Delta_{\mbox{\tiny matter}} \tilde N}
\ee
The difference is that besides one inhomogeneous solution $\Delta N=c$, we will have all the homogeneous ones $\Delta_{\mbox{\tiny matter}}N=0$.  Luckily, we know that $\Delta_{\mbox{\tiny matter}}$ is Fredholm, and thus it has is a finite-dimensional kernel, $\mbox{Ker}(\Delta_{\mbox{\tiny matter}})$. Let us assume that using a Gram-Schimdt algorithm (in the appropriate $L_2$ norm) we can find an orthonormal set that generates  $\mbox{Ker}(\Delta_{\mbox{\tiny matter}})$, we'll call them $\{N_i\}_{i\in I}$, where $i\in \mathbb N$. Note that this is \emph{not} the analogous condition required on the invertible case of the previous section. There we had to \emph{prove} that it was possible to choose $\mean {N_0}=1$. Here, we are using $\mean{N_i, N_j}=\delta_{ij}$, which requires no proof since we are assuming we are in a finite dimensional subspace of a Hilbert space. The reason for this departure will become clear soon. 

 So any solution to the inhomogeneous equation has an added ambiguity given by $\alpha^i N_i$, where $\alpha^i\in \mathbb R$, and assuming $I$ has $n$ elements, the space of solution of \eqref{equ:Delta_constr_matter} is $n+1$ dimensional. We note furthermore that since the solutions $N_i$ depend on the point in phase space we are at, so does the dimension $n$. Lastly, we note that the different $N_i$'s are linearly independent, and of course never identically zero as functionals of $(g,\pi)$: $N_i(g,\pi)\not\equiv 0$. 

\subsubsection{Separating out the purely second class constraints}
Again, we have \eqref {equ:def}
\begin{equation}{ \overline{\mathcal{T}S}(x)}:\Gamma\times T^*(C^\infty(M)/\mathcal{V})\times \Gamma_{\mbox{\tiny matter}} \rightarrow C^\infty(M),\end{equation}  
where $\Gamma_{\mbox{\tiny matter}}$ denotes the phase space of the matter fields. 
Thus, following \eqref{equ:tangent}:
   \be\delta_{\mathcal{C}}\overline{{\mathcal {T}_\phi}S}(g_0,\pi_0,\pi_\phi^0,\psi^A_0,\pi_A^0)_{|\phi=0}: {T}_0(C^\infty(M)/\mathcal{V})\rightarrow C^\infty(M)\ee
and finally from \eqref{equ:tangent2}
 \be(\delta_{\mathcal{C}} \overline{\mathcal {T}_\phi S})^*\cdot N= \Delta_{\mbox{\tiny matter}} N-\langle \Delta_{\mbox{\tiny matter}} N\rangle
 \ee
and from the eliipticity we in fact have 
\be \mbox{Im}(\delta_{\mathcal{C}} \overline{\mathcal {T}_\phi S})\bigoplus_{i=0}^n N_i=C^\infty(M)
\ee

Now we construct the tentative operator 
\be L_0(x):=S(x)-\sqrt g\sum_{ i=0}^n S(N_i)
\ee
We can readily check that the 
tangent map $\delta_{\mathcal{C}} {\mathcal {T}_\phi L_0}_{|_\phi=0}$  indeed stays (weakly)  the same, \emph{but} 
 $\langle L_0, \alpha^iN_i\rangle\neq 0$, even if we were to assume the conditions $\mean {N_i}=1$.
At this point it is already clear that it will be hard to recover the original version of Shape Dynamics, as indeed we don't. 

Thus, we are forced to depart the usual construction of Shape Dynamics, and now we use the fact that we have made the set of generators $N_i$ orthonormal. The operator we are to use becomes:
\be L(x):=S(x)-\sqrt g\sum_{ i=0}^n S(N_i)N_i(x)
\ee
Now it becomes clear that $\mean{L,N_j}=0$, and the tangent map $\delta_{\mathcal{C}} {\mathcal {T}_\phi L}_{|\phi=0}$  is still (weakly) the original. Thus
  \begin{equation}({\mathcal{T}_\phi L})_{\pi_\phi=f(x)}:\Gamma\times C^\infty(M)/\mathcal{V}\rightarrow \text{Im}(\delta_{\mathcal{C}} \mathcal {T}_\phi S).
\end{equation}

To finish the proof that  $\delta_{\mathcal{C}}\overline{\mathcal {T}_\phi L} $ is a topological linear isomorphism when restricted to the correct space,  we still need to check whether  the space of conformal factors needs restrictions so that the resulting  operator is injective. A priori, we would expect that we would also have to quotient the space of conformal factors by an $n$-dimensional space. This would mean that there would be an $(n+1)\times(n+1)$ set of generating Hamiltonians. As we will see, surprisingly this is not the case.

\begin{proposition}\label{prop:inv_matter}
$$(\mbox{Im}(\delta_{\mathcal{C}}\overline{\mathcal {T}_\phi S})^*)^\perp=\mbox{Ker}(\delta_{\mathcal{C}}\overline{\mathcal {T}_\phi S})=\mathcal{V}$$
where $\mathcal{V}$ is the space of constant functions. \end{proposition}
The proof we offer is not very clean, and there might be better ways of attaining the same result. In any case, the proposition yields the surprising result that we can still completely solve the second class constraints for the conformal factor $\hat\phi_o$, as in the vacuum case. 

Again we must analyze  $(\mbox{Im}\delta_{\mathcal{C}}\overline{\mathcal {T}_\phi S}^*)^\perp$. To provide some guidance, and also a necessary preliminary,\footnote{This is the second warm-up to the problem. The first one was given in section \ref{sec:tech_vacuum}, see equation \eqref{equ:injectivity1}.} consider the following simplified case:
suppose that $(\delta_{\mathcal{C}}\overline{\mathcal {T}_\phi S})^* $ was of the simpler form $(\delta_{\mathcal{C}}\overline{\mathcal {T}_\phi S})^*\cdot N=  \Delta_{\mbox{\tiny matter}} N$. Of course, $ \Delta_{\mbox{\tiny matter}}$ is self-adjoint, so we know that 
$\mbox{Ker}((\delta_{\mathcal{C}}\overline{\mathcal {T}_\phi S})^*)=\mbox{Ker}(\delta_{\mathcal{C}}\overline{\mathcal {T}_\phi S})
$ and  we already know the complete characterization of the decomposition once we know the kernel. Nonetheless, let us follow through the computations. 
Using the $L_2$ inner product structure we can thus easily parametrize the image of $(\delta_{\mathcal{C}}\overline{\mathcal {T}_\phi S})^*$ which is just $(\mbox{Ker}(\delta_{\mathcal{C}}\overline{\mathcal {T}_\phi S}))^\perp$,  as:
\be\label{equ:image_param}\{f(x)-\sum_i\mean{f,N_i}N_i(x) ~|~f\in C^\infty(M)\}
\ee
Thus we have that $\rho\in \mbox{Ker}(\delta_{\mathcal{C}}\overline{\mathcal {T}_\phi S})$ if and only if, for all $f\in C^\infty(M)$:
\be \langle {\rho, f-\sum_i\mean{f,N_i}N_i}\rangle=0
\ee
By choosing, in the same way as done in \eqref{equ:injectivity1}, for a given point $y\in M$ (for which $\rho(y)\neq 0$),   $f_y(x):=\rho(x)\delta(x,y)$, we get:
\be\rho(y)-\sum_iN_i(y)\mean{\rho,N_i}=0
\ee
which is a consistent equation solved by $\rho(x)=\sum_i a^iN_i(x)$, as expected. 

Now, let us move forward to the full problem. This time we are trying to find the space orthogonal to 
\be\label{equ:image_param1}\{\Delta_{\mbox{\tiny matter}}N(x)-\mean{\Delta_{\mbox{\tiny matter}}N} ~|~N\in C^\infty(M)\}
\ee
We know that we can further parametrize this by substituting 
$$\Delta_{\mbox{\tiny matter}}N(x)\rightarrow f(x)-\sum_i\mean{f,N_i}N_i(x)$$
using self-adjointness and knowledge of the kernel of $\Delta_{\mbox{\tiny matter}}$.  We thus have that the condition for $\rho$ to be in the kernel of $\delta_{\mathcal{C}}\overline{\mathcal {T}_\phi S}$ is that, for all $f\in C^\infty(M)$:
\be\label{equ:full} \mean{f,\rho}-\sum_i\mean{f,N_i}\mean{\rho, N_i}-\mean{f}\mean{\rho}V+\sum_i\mean{f,N_i}\mean{N_i}\mean{\rho}V=0
\ee
Contrary to what we might have supposed, $\rho=\sum_ia^iN_i$ is not a solution for whichever set of $a_i$'s we choose,  since then  \eqref{equ:full} is equivalent to 
\be\label{equ:f_kernel} \mean{f}-\sum_i\mean{f,N_i}\mean{N_i}=0
\ee  
and $\{N_i\}$ only forms a basis of a finite-dimensional subspace (thus $\sum_i\mean {N_i,\cdot}N_i $ is not a decomposition of the identity). We can also easily see that for $k=\mbox{const}$, $\rho(x)=k$ is a solution of \eqref{equ:full}, and thus for $\rho(x)=h(x)+k$, $h(x)$ has to separately be a solution.

Finally, choose in \eqref{equ:full} $f(x)=\rho(x)\delta(x,y)$, as before. After a tiny bit of algebra we get:
\be \rho(x)=\mean{\rho}+N_i(x)(\mean{\rho, N_i}-\mean{N_i}\mean{\rho}V)
\ee
 Of course, this implies that $\rho(x)=h(x)+k$, where $h(x)=\sum_i a^iN_i(x)$ for some choice of $a^i\in \mathbb R$. As we saw, $h(x)$ must be separately a solution, but from \eqref{equ:f_kernel}, we know that it isn't. Therefore the only solution is $a^i\equiv 0$, thus $\rho(x)=k$ . ~~$\square$

This allows us to find the first and (purely) second class constraints, a necessary step for our construction. From here we can know what is the equation we have to explicitly solve (i.e. the purely second class constraints). Hence even if we are to use the York-\'O'Murchadha, Leray-Schauder method, we must come this far to know \emph{what} is the equation we would like to apply it to. That said, we have been unable to work out how said method might work for the second class constraint found here.

 By the canonical transformation properties of $\mathcal{T}_\phi$, one can extend this construction to arbitrary $\phi$. We now have
\begin{proposition}\label{prop_matter}
At the constraint surface, the linear map given by 
$$\delta_{\mathcal{C}}\overline{\mathcal {T}_\phi L}(x):T_0(C^\infty(M)/\mathcal{V})\rightarrow \text{Im}(\delta_{\mathcal{C}}\overline{ \mathcal {T}_\phi S})\simeq C^\infty(M)/\bigoplus_{i=0}^n N_i$$
where 
\be \label{equ:pure_2nd}
{\mathcal {T}_\phi L}(x)=\mathcal {T}_\phi S(x)-\sum_{i=0}^n \mathcal{T}_\phi( S(N_i)N_i(x))\sqrt{g} 
\ee
 is a toplinear isomorphism.
  \end{proposition}
  We have shown that it is a linear continuous bijection, and hence a topological linear isomorphism. $\square$.  

\subsection{The reduced system}
Proposition \ref{prop_matter}  tells us what part of the constraint is ``purely second class" and thus what part of the constraint should be explicitly solved for. Again we have the Dirac bracket  and  use the implicit function theorem for Banach spaces for the function 
$\overline{\mathcal {T}_\phi L}$ around the constraint surface to assert  that
\begin{theorem}\label{prop}
 There exists a unique $\hat\phi_0:U\subset(\Gamma\times \Gamma_{\mbox{\tiny matter}})\rightarrow V\subset (C^\infty(M)/\mathcal{V})$,  such that the level surface formed from ${\mathcal {T}_\phi L}=0$ is given by  \end{theorem}
 $$\{(g_{ij},\pi^{ij},\psi^A,\pi_ A,
 \hat\phi_0[g_{ij},\pi^{ij}],\pi_\phi)\in U\times (\Gamma_\phi)_{|_V}\}.
$$ In other words, given any point on the constraint surface,  the solution to ${{\mathcal {T}_\phi L}}(g,\pi,\phi,\pi_\phi)=0$ for all $(g,\pi,\psi^A,\pi_ A,\pi_\phi)$ on a given open set around this point is obtained   by setting $\phi=\phi_0$. However, even though we still have that for each point $\delta_{\mathcal{C}}\overline{ \mathcal {T}_\phi L}$ is surjective, the level surface formed by $({\mathcal {T}_\phi L})^{-1}(0)$ may not be regular, since the number of linearly independent solutions $N_i$ may change from point to point in phase space. This fact will not be dealt with here.

 We also note that one of the initial fears of breaking the bound, namely that there would be a different $\phi_o$ for each homogeneous solution, is not realized, since $\phi_o$  comes from the solution of $\mathcal {T}_\phi L=0$, and not from the ``lapse fixing equation". The local implication that there might be different $\phi_o$ functionals around different points $(g,\pi,\psi^A,\pi_A)$ also does not concern us here. 

Finally, we deal with the reduced system. First, note that for none of the solutions $N_i$, we have that  $N_i\equiv 0$, and the $N_i$ are furthermore linearly independent. This entails that if we have a constraint of the form $\sum_i a^iN_i=0$, we can only satisfy it (remember the constraints must be satisfied for all $x\in M$) by setting $a^i=0$. Combining this with the result that $\mathcal{T}_\phi f(g,\pi,\pi^A,\psi_A)(x)\equiv 0 $ if and only if $f(g,\pi,\pi^A,\psi_A)(x)\equiv 0$ (see the appendix in \cite{Gomes:thesis}), we have that the reduced constraints 
\be \sum_{i=0}^n \mathcal{T}_{\phi_o}( S(N_i)N_i(x)) 
\ee
are equivalent to the $n+1$ independent constraints:
\be \{\mathcal{T}_{\phi_o}( S(N_i))~|~{i=0,\dots n}\}
\ee
This resolves a possible paradox with the set of leftover physical degrees of freedom of the theory, which would have been enlarged if there was no such simplification of the constraints. We note that with this new method, which reduces to the vacuum case, previous considerations about $N_0$ (such as the possibility of choosing it such that $\mean {N_0}=1$) are turned obsolete. 

 Finally, we arrive at the generalized Shape Dynamics total Hamiltonian:
\begin{equation}\label{equ:total_hamiltonian_SD}
  H_{\text{SD }}=\alpha^i\mathcal {T}_{\phi_o}(S(N_i))+\int d^3x \left(\pi^{ab}\mathcal (L_\xi g)_{ab} + \pi^A(\mathcal L_\xi \psi)_A+e^{k\hat\phi_o} G^\alpha(g_{ab},\psi_A)\lambda_\alpha\sqrt g\right).
\end{equation}
where we have used the decoupling of the Gauss constraints related to the extra fields, $\mathcal {T}_{\phi}G^\alpha=e^{m\hat\phi} G^\alpha$  which appears for some $m$ for all fields studied (for example, for Yang-Mills $m=6$). 

We take the opportunity here to note the important fact that this theory is \emph{not the same as the original SD}, in the sense that it does not reduce to SD even when there are no non-trivial homogeneous solutions to $\Delta _{\mbox{\tiny matter}}$. What is different is the form of ${\mathcal {T}_\phi L}(x)$ given in \eqref{equ:pure_2nd}. This does not reduce to $\mathcal {T}_\phi S(x)- \mathcal{T}_\phi( S(N_0))\sqrt{g}$, the original case realized for matter obeying the bound. Thus, although the form of the global Hamiltonians is the same, $\{\mathcal{T}_{\phi_o}( S(N_i))$, the functional $\phi_o$ might be different, even if the bound is satisfied.

In any case, it is a trivial exercise to check that the non-zero part of the first-class constraint algebra is given by:
\begin{eqnarray}
\{H^a(\eta_a),H^b(\xi_b)\}&=&H^a([\vec\xi,\vec\eta]_a)\nonumber\\
\label{equ:constraintAlgebra}
\{H^a(\xi_a),D(\rho))\}&=&D(\mathcal{L}_\xi\rho)\\
\{\alpha^i\mathcal {T}_{\phi_o}(S(N_i)),\beta^i\mathcal {T}_{\phi_o}(S(N_i))\}&=&\alpha^i\beta^jH^a(N_j\nabla _a N_i-N_i\nabla _a N_j)\\
\end{eqnarray}

\section{Conclusion}
The present work considers the question put forward previously in work coupling different fields to Shape Dynamics \cite{Gomes:2011au, Gomes:loops}. Namely, it was found that for some types of couplings, including the mere inclusion of a cosmological constant, certain bounds on the previous construction of SD were imposed. In usual GR language, these bounds ensured that there existed a unique lapse propagating a CMC condition, and were technically  implied by the triviality of  the kernel of a given differential operator, called here $\Delta_{\mbox{\tiny matter}} $.  Thus not only for completeness of the theory, but also if we would like to make statements about cosmology in SD, it was seen as important that we sorted out such limitations in the theory, which is what this paper was aimed at.
 
Using the same method of ``excising"  kernels and co-kernels to obtain isomorphisms as brackets of ``purely second class constraints", a method  fully described  in section 4.3  of \cite{Gomes:thesis}, we attempted to extend previous results of standard SD beyond such bounds. Surprisingly, we found a theory slightly different than SD, even when the bounds are respected. In such case the distinction will be manifested  only by the appearance of a different scalar functional $\phi_o[g,\pi,\psi^A,\pi_A](x)$, appearing in a global Hamiltonian of the same form as the original. The other constraints and the constraint algebra are unchanged. 
The general theory found here still exists as a conformally invariant theory, possessing all the advantages of the original version,  but  now it may have also a finite number of weakly commuting global Hamiltonians. The number of Hamiltonians is given by $n+1$, where $n$ is the dimension of the kernel of  $\Delta_{\mbox{\tiny matter}} $, and yields the number of different lapses propagating a CMC condition.

These results also serve to make more distinct the role developed by $\phi_o$ and the ``lapse solutions" $N_i$,  which were previously thought to be 1-1, or correspondent. Before these results we believed that for each homogeneous solution of 
$\Delta_{\mbox{\tiny matter}} $   there would also exist a corresponding different solution of the second class constraints for the Stuckelberg field $\phi$. This would yield a space of $(n+1)\times(n+1)$ additional constraints which might have turned out to be inconsistent. Had there been different reductions, i.e. different solutions for $\phi$,  interpretation of the result would have been extremely difficult, since we would have had several generators of evolution which could not have been shown to commute. To be more explicit, departing from $\{\mathcal{T}_\phi f, \mathcal{T}_\phi h\}=\mathcal{T}_\phi \{f,h\}$, for a single reduction the commutator will be simple to compute, but if we had two reductions $\phi_1, \phi_2$ it becomes unclear what the result might have been.  However, the  reduction process was  shown to be unique, and these fears were not realized.  This surprising result was attained in proposition \ref{prop:inv_matter}.  This work gives us the picture of a general theory of Shape Dynamics, always with a finite number of generators of time evolution, with a related freedom to continuously shuffle between them. 

We would also like to highlight the use of the implicit function theorem as a purely technical tool. Coupled to canonical transformations, it is able to yield local solvability of equations of Lichnerowicz-York type from Poisson brackets in the original phase space $\Gamma\times\Gamma_{\mbox{\tiny matter}}$. This is a very different approach than Leray-Schauder theory  of non-linear differential equations polynomial in $\phi$, employed by York-\'O'Murchadha \cite{York-Niall} to solve a ``conformalized" scalar constraint. The present  method yields a simple, linear criterion for local solvability of such equations.  As far as the author knows, this technique has not been previously employed.
 More importantly, to find the equations one needs to solve for, one must separate out the ``purely" second class constraints. I.e. those that we would like to solve as definitions of the auxiliary variables.This of course has already happened in the vacuum case, whence we arrived at the ``non-homogeneous" version of the LY equation: $k[g,\pi]=\mathcal{T}_\phi S(x)$, where $k[g,\pi]$ was a spatial constant. It is only from this point on that  one can even discuss using the York-\'O'Murchadha  method.  Thus \emph{even to formulate} the equation one wants to solve in terms of the $\phi$ field (e.g.:  $k[g,\pi]=\mathcal{T}_\phi S(x)$ )  one must determine a split of the original scalar constraint such that one of the parts has invertible Poisson brackets with the CMC constraint. This is a necessary technical  point that our dynamical study must arrive at in any case, and from here on the use of the implicit function theorem requires no extra work. 

In connection to this, let  us mention a  practical difference between the generic case presented here and the vacuum case. In the vacuum case, the fact that we could find the solution $\phi_o$ by use of the equation $k[g,\pi]=\mathcal{T}_\phi S(x)$, where $k[g,\pi]$ was a spatial constant, was due to the fact that the purely second class part of $\mathcal{T}_\phi S$ was of the form $\mathcal{T}_\phi S(x)-\mathcal{T}_\phi S(N_0)$. Now the correspondent second class part, given by equation \eqref{equ:pure_2nd}, is not nearly as simple, and the rhs is spatially  dependent. In this case  the York-\'O'Murchadha method does not yield an alternative method of proof, it is only through the implicit function theorem that we can find the right theory. 

Regarding actually finding solutions, in the general case then one has no choice \emph{but to find the individual solutions} $N_i$,  plug them in and work out solutions of $\phi_o$.  Of course, it remains true that if $(g,\pi)$ belong to a physical ADM solution $S(g,\pi)=0$, then we still obtain $\phi_o=0$. Thus a useful approach to find non-trivial solutions of SD is, as in the vacuum case, to perform a perturbative analysis around a given ADM spacetime. 

\section*{Acknowledgements}

The author would like to thank Steve Carlip, for urging us to consider seriously the consequences of breaking the bound for Shape Dynamics, and all related interpretative questions of commuting different evolutions. Also Tim Koslowsk, Sean Gryb, Flavio Mercati and Julian Barbour,  for numerous and ongoing conversations about SD. 
 This work was supported in part by the U.S.
Department of Energy under grant DE-FG02-91ER40674.

\end{document}